# Charming Charm, Beautiful Bottom, and Quark-Gluon Plasma in the Large Hadron Collider Era

*Santosh K. Das[1] and Raghunath Sahoo[2]*

*Abstract:* After a few microseconds of the creation of our Universe through the Big Bang, the primordial matter was believed to be a soup of the fundamental constituents of matter – quarks and gluons. This is expected to be created in the laboratory by colliding heavy nuclei at ultra-relativistic speeds. A plasma of quarks and gluons, called Quark-Gluon Plasma (QGP) can be created at the energy and luminosity frontiers in the Relativistic Heavy Ion Collider (RHIC), at Brookhaven National Laboratory, New York, USA, and the Large Hadron Collider (LHC) at CERN, Geneva, Switzerland. Heavy quarks, namely the charm and bottom quarks, are considered as novel probes to characterize QGP, and hence the produced Quantum Chromodynamics (QCD) matter. Heavy quark transport coefficients play a significant role in understanding the properties of QGP. Experimental measurements of nuclear suppression factor and elliptic flow can constrain the heavy quark transport coefficients, which are key ingredients for phenomenological studies, and they help to disentangle different energy loss mechanisms. We give a general perspective of the heavy quark drag and diffusion coefficients in QGP and discuss their potentials as probes to disentangle different hadronization mechanisms, as well as to probe the initial electromagnetic fields produced in non-central heavy-ion collisions. Experimental perspectives on future measurements are discussed with special emphasis on heavy-flavors as next-generation probes in view of new technological developments.

*Key Words: Big Bang, quark-gluon plasma, heavy-ion collisions, heavy flavor*

[1] Santosh K. Das is at Indian Institute of Technology Goa, India, Email: santosh@iitgoa.ac.in
[2] Raghunath Sahoo, FInstP is at Indian Institute of Technology Indore, India, and CERN Scientific Associate, CERN, Geneva, Switzerland. Email: Raghunath.Sahoo@cern.ch (Corresponding Author)

The second half of the twentieth century was a tremendous success in particle physics through the discovery of the Quark Model of hadrons by Murray Gell-Mann and George Zweig, the Standard Model of Particle Physics by Glashow, Salam, and Weinberg (and many others) through the unification of fundamental forces. The endeavours in basic science, while looking for the fundamental constituents of matter, have contributed immensely to the developments in particle detection and accelerator technologies, which have given direct and indirect social benefits of colossal nature. As the present-day understanding of constituents of matter, we have 6-quarks, 6-leptons, their antiparticles, and the force carriers. However, among these, we only encounter the light quarks – up and down, and electrons in normal nuclear matter. Other heavy particles are created in high-energy interactions in cosmic rays and particle accelerators. Although the free existence of these fundamental particles like quarks and the force carrier of strong interaction- the gluons is not observed because of the confinement property of strong interaction, the quest to create a matter with partonic degrees of freedom has been a frontier of research for decades. For illustration, Fig. 1 shows the confinement of quarks inside hadrons and quark matter consisting of deconfined colored quarks and gluons.

Theoretical calculations based on lattice QCD (lQCD) predict that hadronic matter at high temperatures and high energy densities melts into a deconfined state of quarks and gluons - called Quark-Gluon Plasma (QGP)[1]. In this state, hadrons like protons and neutrons, which form the normal matter around us, lose their identity, and their constituents -- quarks and gluons can move freely in this plasma. The major goals of the ongoing heavy-ion collision programs at Relativistic Heavy Ion Collider (RHIC) at the Brookhaven National Laboratory, USA, and the Large Hadron Collider (LHC) at CERN, Geneva, Switzerland at Giga and Terra electron volt energies are to create and characterize QGP in the laboratory. This state of matter is found to behave like a nearly perfect fluid having a remarkably small value of shear viscosity to entropy density ratio. QGP is believed to be the primordial matter formed at the infancy of our universe just after a few microseconds of the creation of the universe through the Big Bang. This primordial matter of quarks and gluons, because of its large concentration of energy and high temperature, expands with time (like an expanding fireball) through various complex processes. These processes include formation of composite hadrons out of quarks and gluons, called hadronization (corresponding to a system temperature, which marks the quark-hadron phase transition: deconfinement transition); chemical freezeout, where the particle abundances are frozen (inelastic collisions stop) and kinetic freeze-out (ceasing of elastic collisions) followed by free streaming of the secondary particles to reach the detectors sitting all around the interaction point of the hadronic/nuclear collisions in the experimental area. The study of the properties of QGP is a field of high contemporary interest and remarkable progress has been made in the last decades towards the understanding of the properties of strongly interacting Quark-Gluon Plasma. In this article, we shall focus on heavy flavor particles as a powerful probe of this primordial matter and their enriched measurements at the LHC energies given the high luminosity and detector upgrade programs with high-end technologies.

A pictorial representation of three generations of quarks and their masses is given in Fig. 2 for illustration. The masses of charm and bottom quarks are higher than up, down, and strange quarks, which are taken as light quarks (flavors). Although charm, bottom and top are considered as heavy-quarks, because of the huge mass of top quark and its low production cross-section, it is not considered for the present discussions.

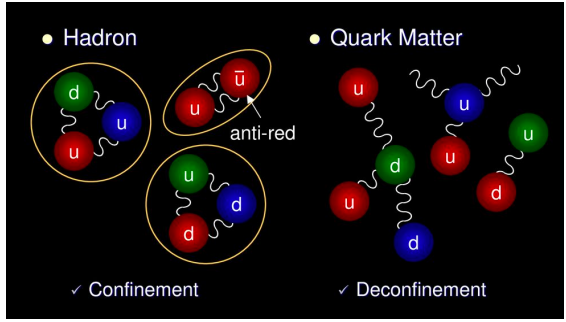

**Fig.1:** Pictorial representation of confinement of quarks inside hadrons (proton, neutron etc.) and the deconfined quark matter called quark-gluon plasma. Figure courtesy[28].

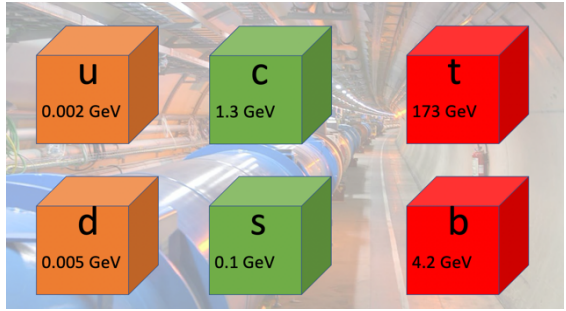

**Fig.2:** Pictorial representation of three generations of quarks and their masses.

Heavy quarks[2,3,4], mainly charm and bottom quarks, because of their large masses, are considered as novel probes to characterize the properties of the QGP phase. These heavy quarks are created in hard processes which are accessible to perturbative QCD (pQCD) calculations. Therefore, their initial distribution is theoretically known and could be verified by experiments. They are produced in the early stages of the collisions and witness the entire space-time evolution of the created fireball and can act as an effective probe of the created QCD matter. Heavy quarks (HQs), as colored objects, interact strongly with the plasma constituents and their momentum spectra are strongly modified. This is expressed by the nuclear suppression factor[5,6], $R_{AA}(p_T)$, the ratio of the measured heavy quarks momentum spectra (yields) in nucleus-nucleus to proton-proton collisions, rescaled by the number of binary collisions. If there were no interactions with the QGP (as well as no cold nuclear matter effects like shadowing) one would indeed find $R_{AA} = 1$. Another observable is the elliptic flow[7,8], $v_2(p_T) = <\cos(2\Phi)>$ of heavy-quarks. Here $\Phi$ is the azimuthal angle of emission of the secondary particles in hadronic/nuclear collisions. It is a measure of the anisotropy in the angular distribution of the produced heavy mesons as a response to the initial anisotropy in coordinate space in non-central nucleus-nucleus collisions, transferred by the bulk to the heavy quarks.

The relaxation time ($\tau_{HQ}$) of a heavy-quark of mass, M at a temperature T is larger than the relaxation time of light partons by a factor of M/T (> 1). In other words, the light quarks and the gluons get thermalized faster than the HQs. The propagation of HQs through QGP can therefore be treated as the interactions between equilibrium and non-equilibrium degrees of freedom. The Fokker-Planck (FP) equation provides an appropriate framework for such processes. Several theoretical efforts have been made to study various observables like $R_{AA}$ and $v_2$ of the heavy quarks within the FP and the relativistic Boltzmann approach, which are then confronted with the experimental measurements. The heavy quarks drag ($\gamma$) and momentum diffusion coefficients ($D_p$) are the key inputs, to solve the FP equation to study the momentum evolution of heavy quarks in QGP medium to compute its $R_{AA}$ and $v_2$, which contain the microscopic details mentioning how the heavy quarks interact with the hot QGP medium. Heavy quark drag coefficient is responsible for the heavy quark energy loss and the diffusion coefficient is responsible for its momentum broadening. The drag and diffusion coefficients of the heavy quarks can be computed starting from the heavy quark - light quark scattering matrix element. The heavy quarks drag, and momentum diffusion coefficients are related through the fluctuation-dissipation theorem (FDT), $D_p = M \gamma T$. Here T is the temperature of the thermal bath and M is the HQ mass. The spatial diffusion coefficient, $D_s$, can be calculated from the diffusion coefficients in momentum space in the static limit (p → 0): $D_s = T^2/D_p$. The major aim of all the theoretical studies is to extract the heavy quark spatial diffusion coefficients ($D_s$), which quantify the interaction of heavy quarks with the medium, that is directly related to the thermalization time and can be evaluated using lattice QCD. The $D_s$, $D_s = T/M^* \gamma$ (in the limit of momentum, p → 0), is directly related to the heavy quark drag coefficient γ. Drag coefficient appearing in $D_s$ is expected to scale with the mass, hence, provides a measure of the QCD interaction ideally independent of the mass of the quark. One can estimate the heavy quark thermalization time, $\tau_{th} = \gamma^{-1}$ (p→0) by:

$$\tau_{th} = (M/2\pi T^2)(2\pi T D_s)$$

A simultaneous description of both $R_{AA}(p_T)$ and $v_2(p_T)$ is a top challenge to all the theoretical models and is known as a *"heavy quark puzzle"*. Before the first experimental observations, it was expected that HQ interaction with the QGP could be characterized by means of pQCD, which led to the expectations of a smaller modification of their spectra and a smaller $v_2$. However, the first observations of non-photonic electrons coming from heavy meson decays measured at the highest RHIC energy have shown a surprisingly large modification of their spectra and a quite large $v_2$, indicating strong interactions between HQs and the QGP medium. This is substantially beyond the expectations from pQCD interactions. These observations have triggered several investigations in which non-perturbative effects have been incorporated. In the recent past, a simultaneous study of both the observables, $R_{AA}(p_T)$ and $v_2(p_T)$, received significant attention[10], as it has the potential to constrain the temperature dependence of HQ transport coefficients in QGP medium and disentangle different energy loss mechanisms. Recent studies based on temperature and momentum dependence of transport coefficients and Bayesian model to data analysis have obtained similar conclusions from different viewpoints[11,12].

Interestingly, the origin of the $v_2$ is quite different for the heavy quarks than the bulk matter. The bulk develops its $v_2$ from the combined effect of the initial coordinate space anisotropy created in non-central collisions and the interaction among the quarks and gluons which constitute the bulk. However, heavy quarks develop a substantial part of the $v_2$ due to their interactions with the bulk medium and partly from the light partons as a consequence of hadronization[13]. A coupled study of light and heavy particle flow harmonics in heavy-ion collisions on event-by-event basis is another interesting topic. Experimental measurements in this direction will offer further new insights on heavy quark - bulk in-medium interaction and the temperature dependence of the heavy quark transport coefficients.

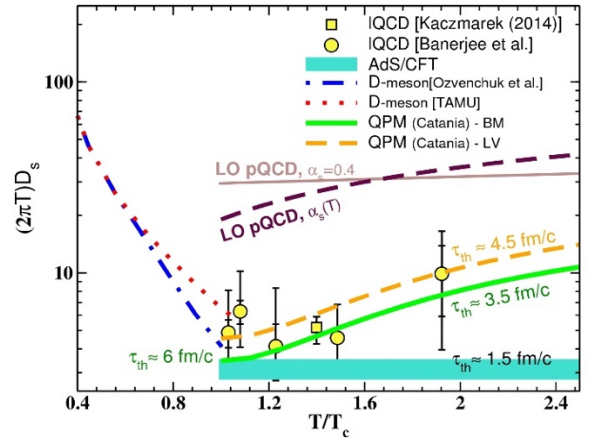

**Fig.3:** Spatial diffusion coefficient as a function of scaled-temperature obtained within different theoretical models to describe experimental data, along with the results from lQCD[9]. Here, $T_c$ is the critical temperature for a quark-hadron transition.

In Fig.3, we show the variation of HQs $D_s$, a standard quantification of the space diffusion coefficient is done in terms of a dimensionless quantity, obtained within different models in comparison with the lQCD results. Being a mass-independent quantity it can provide a general measure of the QCD interaction. Additionally, this quantity can be calculated within the framework of lQCD, as it is directly related to the spectral function. This highlights heavy quarks as a probe of QGP, which have the potential to link the phenomenology constrained from the experimental data to lQCD for studying the transport properties of the hot QCD matter. Energy loss experienced by energetic partons leads to a suppression of final hadrons with high momentum known as jet quenching. At high momenta, one can extract the heavy quark jet quenching parameter, the average transverse momentum broadening squared per unit length, which contains information regarding jet-medium interactions.

Once the temperature of the system goes below the quark-hadron transition temperature heavy quarks convert to hadrons, through a process called hadronization. Hadronization dynamics play an important role in determining the final momentum spectra and therefore, the $R_{AA}$ and $v_2$ in both the light and heavy quark sectors. In particular, for heavy quarks, it is generally expected that a coalescence mechanism[13] is in action especially in the low and intermediate momentum. Heavy baryon to heavy meson ratios, ($\Lambda_c/D$ and $\Lambda_b/B$), are considered fundamental for the understanding of in-medium hadronization[14,15]. Furthermore, the heavy baryon to meson ratio can serve as a tool to disentangle different hadronization mechanisms[14,16].

With the advent of very high-energy accelerators like those are at the RHIC and LHC, we are able to create extremely strong magnetic fields in non-central heavy-ion collisions. The estimated values of the initial magnetic field strengths ($10^{19}$ Gauss) are several orders of magnitude higher than the values predicted at the surface of magnetars. This is the highest ever produced magnetic field on the earth created by human beings. But the magnetic field strength rapidly decreases as two ions recede from each other. However, this remains much stronger than the critical field (Schwinger field) during the lifetime of the created fireball. Since the HQs are produced at the early stages of heavy-ion collisions, their dynamics will be affected by such a strong magnetic field[17] and they will be able to retain these effects till their detection as open heavy-flavors like D mesons in experiments. HQs directed flow $v_1$, is identified as a novel observable to probe the initial electromagnetic field produced in high-energy non-central heavy-ion collisions. The sign of the directed flow for a charged particle will be opposite to its anti-particle mainly due to the response of opposite charges to the electromagnetic field. The recent LHC measurement[18], along with the RHIC findings[19], on the D-meson $v_1$, give the indications of the strong electromagnetic field produced in high energy heavy-ion collisions. It can be mentioned that the magnitude of the heavy quark directed flow at LHC energy is about 1000 times larger than that of light quarks. This is mainly because HQs are produced in the early stages of the collisions, hence, witness the peak of the electromagnetic field. Furthermore, the HQs, due to their large relaxation time in comparison to light quarks, are capable of retaining the memory of the initial non-equilibrium dynamics more effectively. Hence, they carry a stronger signal of the early electromagnetic fields. Further, the momentum anisotropy present in the QGP medium can induce

Chromo-Weibel[3] instability[20]. The impact of the anisotropy on heavy quark transport is quite significant as compared to the case while HQs are moving in an isotropic QGP medium[21]. Non-perturbative effects on HQ transport coefficients near the $T_c$ play a significant role. To describe both the $R_{AA}$ and $v_2$ simultaneously, the inclusion of non-perturbative effects is essential[10].

As mentioned before, heavy quarks are produced in the very early stages of heavy-ion collisions due to their large masses. Hence, it can probe the pre-equilibrium phase, the phase produced before the formation of QGP. The effect of the pre-equilibrium phase might be more significant for low-energy heavy-ion collisions: for example, in the case of Au+Au collisions at RHIC energy, results based on event simulation show that equilibration is achieved approximately within 1 fm/c. Whereas, the lifetime of the QGP phase turns out to be about 5 fm/c. Hence the lifetime of the out-of-equilibrium phase is approximately 20% of the total lifetime of the QGP. Currently, heavy quarks are also used as a tool to probe the Glasma[22], strong gluon field produced in the early stages of high-energy heavy-ion collisions.

On the experimental front, there have been several measurements on heavy-flavor spectra, nuclear modification factor, elliptic and directed flow in proton-proton and heavy-ion collisions. However, going to lower transverse momentum and dealing with signal-to-background ratio has been a challenge so far, because of technological limitations. Contrary, the future is more exciting, because ALICE, which is a dedicated experiment for studying QGP, has entered into detector upgrades with Gas Electron Multiplier-based Time Projection Chamber to deal with high-multiplicity and high-luminosity, the Inner Tracking System (ITS) with lower material budget to enhance the signal-to-background ratio and detection efficiency at lower transverse momenta. As the lifetimes of heavy-flavors are very less, their decay lengths (l = c$\tau$, c is the speed of light in vacuum and $\tau$ is the lifetime of the particle in its local rest frame) are small and hence, they decay very close to the interaction vertex by creating a secondary vertex. This necessitates tracking devices nearer to the interaction point with less material budget to reduce the detection background. To have a better grasp, for instance, the innermost silicon tracker of the CMS experiment at the LHC is at 33 mm distance from the interaction point and the ALICE ITS is 39 mm from the interaction point. Because of the high mass, all the heavy-flavors decay before reaching the detectors, and their detection are based on the invariant mass reconstruction, where one encounters high background from the combinatorics (daughters from other sources and material interactions). This makes the experimental studies of heavy-flavors a highly complex endeavour.

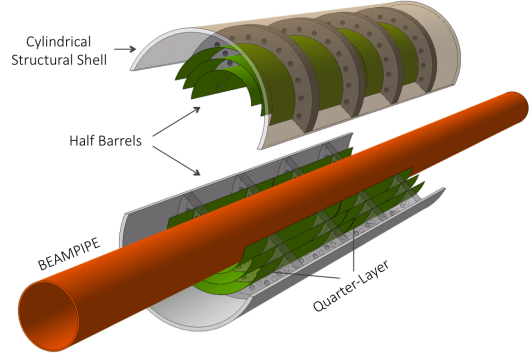

**Fig.4:** Layout of the planned ALICE ITS3 Inner Barrel to be installed during Long Shutdown-3 (LS3) at the LHC. The figure shows the two half-barrels mounted around the beam pipe[23].

---

3 The Weibel instability is a plasma instability present in the electromagnetic plasmas, which arises when the momentum distribution of the charged particles is anisotropic. The non-abelian analogue of the Weibel instability is termed as Chromo-Weibel instability, which also follows from the anisotropy in the momentum space.

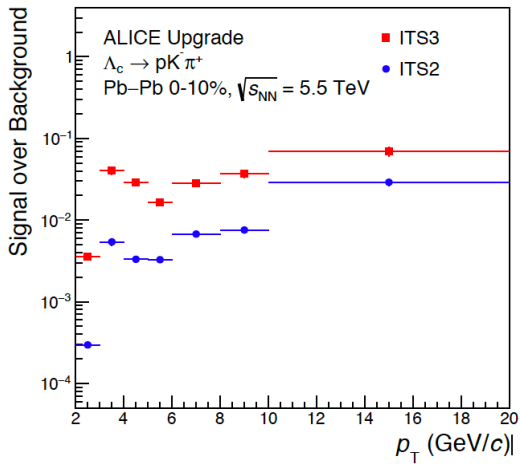

**Fig.5:** $\Lambda^+_c \to pK^-p^+$ in central Pb–Pb collisions at $\sqrt{s_{NN}}$ = 5.5 TeV ($L_{int}$ =10 nb$^{-1}$): signal-to-background ratio (S/B) as a function of $p_T$ (transverse momentum)[23]. ITS3 shows a significant enhancement of S/B for $\Lambda^+_c$ (the lightest charmed baryon) over ITS2 of ALICE.

To overcome these problems, the LHC upgrade plans with the next level of technology, are highly encouraging. ALICE – the dedicated experiment at the LHC to study the primordial matter – QGP has planned for the installation of ITS3 (the 3$^{rd}$ generation Inner Tracking System) in LHC Long Shutdown-3 (LS3). This will have a novel vertex detector consisting of three curved wafer-scale ultra-thin silicon monolithic pixel sensors arranged in perfectly cylindrical layers, which is shown in Fig.4. This will have an unprecedented low material budget of 0.05% X0[4] per layer, with the innermost ITS layer positioned at only 18 mm radial distance from the interaction point (IP)[23,24]. This upgrade is expected to overcome the present limitations in terms of proper identification of secondary vertices, efficiency at low-$p_T$ and dealing with signal-to-background ratio to a greater extent. To have a quantitative understanding, Fig. 5 shows an expected order of magnitude better signal-to-background ratio in the detection of $\Lambda^+_c$, as compared to the ITS2 (the ITS which is being installed during the present ongoing Long Shutdown-2 (LS2)). This will help with the higher detection efficiency of particles containing heavy quarks, opening a new domain of QCD studies. In view of the LHC plan for dedicated high-multiplicity pp runs, the future directions of research and the search for QGP droplets in high-multiplicity proton-proton collisions[25] stay exciting with HQs as potential probes.

Beyond the energy reach of the present Large Hadron Collider, the planned Future Circular Collider (FCC) at CERN is expected to collide proton on proton at a center-of-mass energy of 100 TeV and heavy-ion collisions at 39 TeV per nucleon with a circumference of 100 km, as compared to 27 km of the LHC[26]. This new energy and luminosity frontier would facilitate to study the signals of the early universe in proton-proton and heavy-ion collisions with a variety of new observations with special emphasis on heavy-flavors.

---

[4] X0 is called radiation length and is the mean path length (in gm.cm$^{-2}$) required to reduce the energy of an electron by the factor 1/e (energy loss through Bremsstrahlung radiation– depends on atomic number and atomic mass of the material).

This article gives a theoretical and experimental status of the characterization of the primordial matter – Quark-Gluon Plasma using heavy-flavors as a potential probe. Future research in this direction with improved technologies and energy/luminosity would help in resolving many theoretical issues and having a better understanding of the QCD matter formed in high-energy collisions.

Readers not familiar with the subject of quark-gluon plasma, are advised to go through Ref.[27].

*Acknowledgements*
We would like to thank Prof. Y. P. Viyogi, INSA Senior Scientist, New Delhi, Prof. Tapan K. Nayak, NISER Bhubaneswar & CERN, Geneva, and Dr. Andrea Dainese, INFN Padova, Italy for careful reading and useful comments on this work